# An Overview of the Risk-based Model of AI Governance

## Veve Fry

## Supervisor Kathryn Henne

# Introduction

Artificial intelligence (AI) has become deeply integrated into human life. The harms that AI systems cause are serious and diverse, and as development and uptake of AI systems increases, the potential for broader and more severe harm increases. As a result, the topic of artificial intelligence governance has become an issue of increasing priority for governments, scholars, and citizens. At the forefront of attempts to govern AI, the European Union's "Artificial Intelligence Act" (the AI Act) has been the first of its kind that attempts to impose a comprehensive structured approach to AI governance. This approach follows a long-term trend toward what is called risk governance, and establishes what has been called a 'risk-based' approach to AI regulation. Other jurisdictions have quickly followed the example of the AI Act in the EU, with Australia, Canada and the United States taking steps to implement a risk-based model of AI governance.

Proponents of the risk-based approach argue that by classifying the risks that AI systems pose, the most serious risks, such as those to human life and fundamental rights, can be mitigated, while AI systems that do not pose serious risks can be free from regulatory burden. In this way, the risk-based model can preserve the benefits that AI provides to society and economy, while the most serious harms are mediated. In this paper, I challenge this narrative, arguing that there are flaws with the proposed risk-based approach that require attention and further research. In the section titled *What is the Risk-based Approach?* I introduce the concept of risk-based regulation using perspectives from governance literature, and the origins of the model as it exists in regulatory systems today. Using the examples of food safety regulations in the European Union and financial regulation by the Australian Prudential Regulation Authority in Australia, I argue that the definition of what exactly constitutes a 'risk-based' approach is not agreed upon by scholars or governance officials, but rather the term 'risk-based' is used to refer to systems that use risk analysis and categorisation to inform regulatory decisions.

In *The AI policy landscape in Europe, North America and Australia* section, I summarise the existing AI policy efforts across these three jurisdictions, with a focus on the EU AI Act and the Australian Department of Industry, Science and Regulation's safe and responsible AI consultation, paying attention to the stated objectives of regulation and their justifications. Based on these policies, in the *Analysis* section of this paper, I level several criticisms of the risk-based model of AI governance, arguing that the construction and calculation of risks they propose reproduces existing inequalities. Drawing on the work of Julia Black,[1] I distinguish the concept of 'risk' from 'harm' and argue that certain AI technologies do not pose 'risks' so much as they present deep uncertainties in which the probability and impact of undesirable outcomes are incalculable. Moreover, I argue that the notion of risk itself, due to its inherent normativity, reproduces dominant and harmful social narratives about whose interests matter the most, and excludes marginalised groups. I conclude with a review of governance scholarship that may provide valuable insights into future analysis of the risk-based model of AI governance and suggest that a focus on responsiveness and the use of multiple regulatory instruments should be considered in the continuing development of the model.

---

[1] Black, Julia, 'The Role of Risk in Regulatory Processes', in Robert Baldwin, Martin Cave, and Martin Lodge (eds), The Oxford Handbook of Regulation (2010; online edn, Oxford Academic, 2 Sept. 2010), https://doi.org/10.1093/oxfordhb/9780199560219.003.0014.

## What is the Risk-based Approach?

Risk-based approaches to regulation have been present in regulation discourse for decades.[2] They became particularly popular throughout the 1990s in Britain, in response to what was seen as a crisis of over-regulation that burdened industry and impeded development.[3] At the time, the risk-based approach was framed by policy makers as a better method for allocating resources to where they were most needed, while allowing industry to continue with development with less cumbersome regulations from the state. The risk-based model remains popular in the contemporary governance and regulation discourse for similar reasons- regulators often have less time and resources than needed to effectively address regulatory problems.[4] In its simplest sense, risk-based regulation offers regulators a way to assess which actors require the most regulation, allowing them to allocate time and resources where they are needed most, while at the same time appeasing industry by allowing lower-risk actors to move more freely. It does so by categorising actors or activities according to the risk that noncompliance poses to the regulator's objectives.[5] The basic idea behind risk-based governance is that once the possible risks have been identified, the regulator will dedicate a proportional amount of resources toward mitigating the relevant risks, thus reducing risks where they are most serious.

Since the early 2000s, risk-based framework has become a popular approach to governance across various industries such as food safety, environmental protection,[6] medicine and biotechnology,[7] and banking and finance, among others. But due to the diversity of industries and contexts in which it is used, a common definition of what a 'risk-based' approach means is difficult to pin down, both in literature and regulatory discourse. Commonalities among risk-based governance models include defining, measuring, and prioritising risk, and then using that assessment of risk to influence governance decisions such as which actors, products or firms should be banned, regulated more closely, or left alone. Moreover, the risk-based approach to governance involves certain important tasks for regulators that invariably accompany the use of risk in regulatory process. Black and Baldwin argue that there are five essential tasks for regulatory authorities that characterise any risk-based approach. The regulator must define to whom or what a risk exists; they define what level of risk is acceptable or tolerable; engage in risk assessment; categorise risk; and link the risk categorisation to regulatory institutions and resources.[8] While risk-based models have evolved in the past decades, these core elements remain fundamental to risk-based approaches to regulation because they are, even when not fully acknowledged by regulatory authorities, always decisions that must be made in the development of a risk-based model. Today, common examples of risk-based regulation can be seen in the food safety sector, in environmental regulation, and financial regulation.

---

[2] OECD, "Risk-Based Regulation," OECD Regulatory Policy Outlook 2021 | OECD iLibrary, 2021, https://www.oecd-ilibrary.org/sites/9d082a11-en/index.html?itemId=%2Fcontent%2Fcomponent%2F9d082a11-en.

[3] Hutter, Bridget M. "The attractions of risk-based regulation: accounting for the emergence of risk ideas in regulation." 1.

[4] OECD, "Risk-Based Regulation," OECD Regulatory Policy Outlook 2021 | OECD iLibrary, 2021, https://www.oecd-ilibrary.org/sites/9d082a11-en/index.html?itemId=%2Fcontent%2Fcomponent%2F9d082a11-en.

[5] Black, Julia, and Robert Baldwin. "Really responsive risk-based regulation." Law & policy 32, no. 2 (2010): 181-213. P. 181.

[6] OECD, "Risk-Based Regulation," OECD Regulatory Policy Outlook 2021 | OECD iLibrary, 2021, https://www.oecd-ilibrary.org/sites/9d082a11-en/index.html?itemId=%2Fcontent%2Fcomponent%2F9d082a11-en.

[7] See Ploug, Thomas, and Søren Holm. "The right to a second opinion on Artificial Intelligence diagnosis—Remedying the inadequacy of a risk-based regulation." Bioethics 37, no. 3 (2023): 303-311.

A salient example of risk-based regulation often appears in the food safety industry. The European Union, for example, follows a risk-based approach to meat safety. The EU General Food Law sets out that "official decisions should be based on risk assessments that are independent, objective, transparent and based on available scientific data,"[9] thereby establishing what can be called a risk-based approach.[10] What this looks like in practice is that meat safety regulations introduced the concept of 'food chain information' into the governance model, which requires abattoirs and farmers to share information about food risk between one another. For example, when an examination of an animal or carcass indicated a 'risk' was present, the animals would be subjected to a further examination procedure. 'Risk,' in this context, is thought of in terms of risk to public health, and is calculated in terms of the presence or absence of certain bacteria that impact consumers' health, for example, *Salmonella Enterica and Escherichia Coli,* which are bacteria that cause food poisoning.[11] In this instance, the European meat safety industries' approach could be said to have defined to whom the risk exists (public health) and a way to measure risk (measuring the presence of bacteria), according to Black and Baldwin's definition. The emergence of the risk-based governance structure in this case is also an important consideration- constructivist approaches to risk-governance scholarship see risk as an epistemological construct that becomes part of a "regulatory rhetoric" legitimating regulatory choices. This might explain why the meat-safety governance structure adopted a risk-based approach to regulation after the highly scandalised outbreaks of 'mad cow disease' (bovine spongiform encephalopathy).[12]

Another example of risk regulation can be seen in the Australian Prudential Regulation Authority's (APRA) approach to regulating the financial sector in Australia. APRA regulates banking, mutual, general insurers, health, and life insurers, friendly societies, and most of the superannuation industry in Australia.[13] APRA began to adopt a risk-based approach to regulation in 1999 and continues to use it today,[14] and, as is often if not always the case with risk-based governance, its aims are to manage risk within the financial system (although APRA makes clear that failures in the system are inevitable- it simply mitigates their probability) while also fostering competition and stability. They do so by employing a so-called "*three lines of defence risk management and assurance model.*"[15] The first line of defence begins at the level of business management who is charged with identifying, assessing, mitigating and monitoring risk, and they are not permitted to abrogate this responsibility.[16] The 'second line of defence' is independent of the first line of defence and comprises "specialist risk management functions,"[17] essentially doing the work of developing systems to identify, assess and manage risk and report it to the Board of APRA and reviewing and challenging risk management systems across the financial sector. The 'third line of defence' comprises functions that provide the Board annually with independent assurance that the risk management framework has been complied with, and every three years with a review of the appropriateness and effectiveness of the risk management framework.[18]

---

[8] Black, Julia, and Robert Baldwin. "Really responsive risk-based regulation." 186.
[9] Blagojevic, Bojan, Truls Nesbakken, Ole Alvseike, Ivar Vågsholm, Dragan Antic, Sophia Johler, Kurt Houf et al. "Drivers, opportunities, and challenges of the European risk-based meat safety assurance system." Food Control 124 (2021): 107870. 3.
[10] See Black, Julia. "Managing regulatory risks and defining the parameters of blame: A focus on the Australian Prudential Regulation Authority." Law & Policy 28, no. 1 (2006): 1-30. P. 4.
[11] Blagojevic, Bojan, Truls Nesbakken, Ole Alvseike, Ivar Vågsholm, Dragan Antic, Sophia Johler, Kurt Houf et al. "Drivers, opportunities, and challenges of the European risk-based meat safety assurance system." P. 3.
[12] Borraz, O., et al. "Why regulators assess risk differently: Regulatory style, business organization, and the varied practice of risk-based food safety inspections across the EU." Regulation & Governance, 16(1), (2022) 274–292.
[13] APRA. "What We Do | APRA," n.d. https://www.apra.gov.au/what-we-do#:~:text=The%20Australian%20Prudential%20Regulation%20Authority,members%20of%20the%20superannuation%20industry.

As the cases of food safety and financial regulation show, approaches to 'risk-based' regulation vary considerably across industries in practice. In governance literature, Julia Black argues,[19] views on the principles of risk-based regulation are diverse- in one sense the most basic principle in a 'risk based approach' to governance is that assessments of risk are used to set standards of conduct. Another conception emphasises its ability to tailor regulatory conduct to be more effective or appropriate, as opposed to simply allocation of resources.[20] Risk-management systems can also be built into a regulatory organisation itself, like in some private-sector companies. Alternatively, risk can be conceptualised institutionally, as arising outside the regulatory organisation and risk is seen as risk to the objectives of the regulator.[21] This last conception, Black argued, was adopted in Britain and Australia and was becoming more popular at the time she was writing. Today, the literature mostly reflects this final conception of the risk-based approach- that risk arises within the industry and exists as a risk to the objectives of the regulatory agency. Still, approaches to managing and calculating industry risks are subject to variation. Even within the food-safety industry in the EU, in countries that exist under the same uniform regulations, approaches to measuring and evaluating risks vary considerably.[22] As such it can be said that there is no single agreed upon set of practices or systems that characterise a 'risk-based' approach. Rather the label 'risk-based' is understood as applying to a broad range of individual governance efforts in different contexts, and is perhaps more useful as an identifier for certain commonalities, as opposed to a discrete category.

---

[14] Black, Julia. "Managing regulatory risks and defining the parameters of blame: A focus on the Australian Prudential Regulation Authority." Law & Policy 28(1) (2006): 1-30. P. 4.
[15] APRA. "Prudential Practice Guide." (2018). WWW.APRA.GOV.AU. https://www.apra.gov.au/sites/default/files/cpg_220_april_2018_version.pdf. P. 6.
[16] Ibid.
[17] APRA. "Prudential Practice Guide." P. 7.
[18] Ibid. P. 8.
[19] Black, Julia, and Robert Baldwin. "Really responsive risk-based regulation." P. 186.

[21] Black, Julia. "Managing regulatory risks and defining the parameters of blame: A focus on the Australian Prudential Regulation Authority." Law & Policy 28, no. 1 (2006): 1-30. P. 4.
[22] Borraz, O., et al. "Why regulators assess risk differently: Regulatory style, business organization, and the varied practice of risk-based food safety inspections across the EU." 274-292.

# The AI policy landscape in Europe, North America and Australia

Today, AI policy regulation is increasing in priority for regulators, states and international bodies. Efforts have been made by state regulators, as well as international organisations and domestic groups, to establish laws that regulate AI, ethics principles to govern it, and to a lesser extent, accountability mechanisms to hold AI actors responsible. One of the most notable endeavours to regulate AI to date has been the European Union's 'Artificial Intelligence Act' (the AI Act).[23] The risk-based approach is a fundamental part of the European Union's approach to AI governance, with the AI Act being the first act of its kind that seeks to limit the use of potential harmful AI technologies.

The official objectives of the AI Act are to: ensure that AI systems placed on the Union market and used are safe and respect existing law on fundamental rights and Union values; ensure legal certainty to facilitate investment and innovation in AI; enhance governance and effective enforcement of existing law on fundamental rights and safety requirements applicable to AI systems; facilitate the development of a single market for lawful, safe and trustworthy AI applications and prevent market fragmentation.[24] The Act emphasises the union's focus on 'high risk' and 'unacceptable risk' applications of AI while also allowing low risk AI systems to go to market without regulatory interference, preserving the economic benefits of AI and preventing market fragmentation. The AI Act is the first act that prohibits the use of certain technologies. The Act does not specify which technologies should be banned by name, rather, it categorises technologies by what they do. AI systems that are banned under the act include AI systems that propose an 'unacceptable risk,' including those;

- deploying subliminal, manipulative or deceptive techniques that cause 'significant harm,'
- exploiting vulnerabilities related to age, disability or socioeconomic circumstances causing 'significant harm,'
- biometric categorisation systems inferring sensitive attributes (race, political opinion, trade union membership, religious or philosophical beliefs, sex life and sexual orientation) except labelling or filtering of lawfully acquired datasets or by law enforcement;
- Using social scoring;
- Assessing the risk of an individual committing crimes based on profiling, except when used to augment human assessments based on 'objective verifiable facts;'
- Compiling facial recognition databases by untargeted scraping of facial images from CCTV or the internet;
- Inferring emotions in workplaces or educational institutions, except for medical or safety reasons;
- Real-time remote biometric identification in public spaces by law enforcement, except when searching for missing persons, preventing substantial threat to life or foreseeable terrorist attack, or identifying suspects in serious crimes.[25]

---

[23] See EU Artificial Intelligence Act. "High-level Summary of the AI Act | EU Artificial Intelligence Act," n.d., https://artificialintelligenceact.eu/high-level-summary/.
[24] Ibid.
[25] EU Artificial Intelligence Act. "High-level Summary of the AI Act | EU Artificial Intelligence Act," n.d., https://artificialintelligenceact.eu/high-level-summary/.

Systems considered to fall under the category of 'unacceptable risk,' when considered to be technologies that go against the EU values, for example, by violating fundamental rights. Below 'unacceptable risk' in the risk hierarchy of the act are 'high risk' AI systems. High risk systems are any AI systems used;

- in critical infrastructure that could put the life and health of citizens at risk;
- education or vocational training that could determine access to education or the professional course of someone's life;
- safety components of products;
- employment and management systems;
- essential private and public services;
- law enforcement that may interfere with people's fundamental rights;
- migration, asylum and border control management;
- administration of justice and democratic processes.

For high-risk systems, the Act will "*set clear requirements for AI systems for high-risk applications; define specific obligations deployers and providers of high-risk AI applications; and require a conformity assessment before a given AI system is put into service or placed on the market.*"[26] Risks that do not fall under unacceptable or high risk categories are considered 'limited' risk, and are subjected to some less restrictive transparency regulations, or 'minimal' risk, and are left unregulated.[27]

The subject of regulation under the AI Act- the actors being regulated, primarily, are the providers of AI systems, meaning the people or company who develop the AI system with the goal of bringing it to market. According to the European Commission,[28] in practice, high-risk AI systems after being developed, will be subjected to the 'conformity assessment' and must comply with 'AI requirements.' Then the AI system must be registered in an EU database of stand-alone AI systems, after which a declaration of conformity will be signed. Then, the product can go to market. After going to market, authorities take charge of surveillance, and the deployers of the product also monitor and report serous incidents and malfunctions to the authorities, and if 'substantial change' happens throughout the AI system's life cycle, the product will once again have to clear the conformity assessment and comply with AI requirements. The conformity assessment is a process of determining whether the requirements set in the Act Chapter III, Section 2, article 9, which are requirements for high-risk systems, have been met.[29] These include the establishment of a 'risk management system' for all high-risk AI systems, that is understood to be a continuous iterative process throughout the product's 'life cycle,' consisting of identifying risks to health, safety or fundamental rights, estimating their likelihood when the product is used as it was intended or under circumstances of 'reasonably foreseeable misuse,' identifying other risks according to the post-market monitoring tool (detailed in article 72), and finally establishing a risk management system for the identified risks.[30] Section 2, Article 9 (4) of the Act also details that these resulting risk management measures will take into account the possible effects of combining the risk management measures in this section *"with a view to minimising risks more effectively while achieving an appropriate balance in implementing the measures to fulfil those requirements."*

---

[26] European Union. Shaping Europe's Digital Future. "AI Act," June 26, 2024. https://digital-strategy.ec.europa.eu/en/policies/regulatory-framework-ai.
[27] EU Artificial Intelligence Act. "High-level Summary of the AI Act | EU Artificial Intelligence Act," n.d., https://artificialintelligenceact.eu/high-level-summary/.
[28] Ibid.
[29] Conformity Assessments Under the AI Act. Securiti. 2024. https://securiti.ai/conformity-assessments-under-the-eu-ai-act/#:~:text=Conformity%20assessment%20is%20the%20process,AI%20systems%20have%20been%20fulfilled. Accessed July 10, 2024.

Australian AI regulation is following the footsteps of the EU AI act. Currently no legislation specifically regarding AI regulation has been passed federally, however the Department of Industry, Science, and Resources (DISR) has opened an inquiry into 'Safe and Responsible Use of AI,' which, in its interim report, concluded that the current Australian regulatory measures on AI are insufficient and a risk-based approach should be adopted.[31] The interim response asserts that many AI technologies pose no risk, or very low risk, some risks are new and require more than just voluntary regulation, and acknowledges that current regularity efforts do not sufficiently address known risks such as discrimination and bias. A risk-based approach, according to the interim response, would help to address the some of the emerging and 'high risk' outcomes from AI that were identified by submissions, including 'technical risks' such as biased training data and resulting unfair outcomes, 'unpredictability and opacity' of AI systems that make it difficult to identify harms and challenge outcomes, 'domain-specific risks' that arise when AI systems interact with existing harms and systems, 'systemic risks' such as so-called 'frontier models' and the use of generative AI systems that produce unforeseen and unpredictable outcomes, and 'unforeseen risks' that will be produced from the speed of development of these systems. The proposed model draws on the definition of high-risk as seen in the EU AI Act, and proposed that the classification of 'high risk' would apply to AI systems whose outcomes were *"systemic, irreversible and perpetual,"*[32] such as as systems used in critical infrastructure, medical devices, systems determining access to education or jobs, used in law enforcement, border control and administration of justice, biometric identification, and emotion recognition. The report asserts that a risk-based approach would be appropriate because where such 'high risk' applications of AI systems will require regulation, there are applications in which AI poses vey little risk and should not be subjected to additional guardrails, such as AI systems for sorting parcels or optimising internal business practices.[33] The benefits of the risk-based approach in the report focused on its purported ability to allow such low-risk applications of AI to be used and developed without the burden of regulation, including minimising costs for businesses that do not use 'high risk' AI, balancing the costs of regulatory burden with the value of risk reduction, and regulatory responsiveness to developments in AI, but also noted that downfalls of the risk-based model included the potential that the framework would not accurately quantify and predict risks, that certain context-specific risks would not be captured by the model, and that 'unpredictable risks' such as those resulting from frontier models, would not be considered.[34]

---

[30] EU AI Act. Chapter III, Section II, article 9. https://artificialintelligenceact.eu/article/9/
[31] Department of Industry Science and Resources (2024). "The Australian Government's interim response to safe and responsible AI consultation, Department of Industry Science and Resources." Available at: https://www.industry.gov.au/news/australian-governments-interim-response-safe-and-responsible-ai-consultation (Accessed: 15 June 2024).
[32] Ibid, p. 14.
[33] Ibid, p. 15.
[34] Ibid, p. 13.

Overall the interim report suggests the use of the risk-based model for AI regulation in Australia would be appropriate but does touch on some of the potential downfalls. Other regulation in Australia has been introduced that reflects this.

The DISR has also introduced "Australia's AI Ethics Principles" as part of "Australia's Artificial Intelligence Ethics Framework," which is a set of 8 voluntary principles that can be adopted by industry, with the goals of helping to achieve "safer, more reliable and fairer outcomes for all Australians; reducing the risk of negative impact on those affected by AI applications; and businesses and governments to practice the highest ethical standards when designing, developing and implementing AI."[35] The principles include "human, societal and environmental wellbeing; human-centred values; fairness; privacy protection and security; reliability and safety; transparency and explainability; contestability; and accountability." According to DISR, by adopting these principles, organisations can build public trust, drive consumer loyalty in AI-enabled services, positively influence outcomes from AI, and ensure all Australians benefit from it.[36] These principles have been used to inform what the DISR has called the "National framework for the assurance of AI in government." This is a policy with the aim of gaining public confidence and trust in the "safe and responsible use of AI by Australia's governments."[37] The framework does not set out specific requirement but rather is meant to establish a foundation for 'assurance' across all aspects of government, with individual jurisdictions developing and adopting their own individual strategies to implement these principles and foster public trust in governmental use of AI systems. As of 2022, New South Wales is the first jurisdiction to have adopted an AI Assurance Framework. In practice, this means that any agency projects exceeding a certain amount of funding will be guided through the process of complying with the NSW mandate to use the AI Ethics Policy and the AI Assessment Framework.[38] The NSW AI Assessment Framework is designed to ensure that AI projects comply with the AI Ethics Principles, and involves "*a systematic process for identifying, documenting and mitigating AI-specific risks,*" involving a self-assessment tool and those completing the self-assessment to rate the residual risk of a project across five ethics principles, with the resulting level of risk determining whether the project should be reviewed, changed, stopped or allowed to continue unchanged.[39]

In North America the approach to regulating AI shows many similarities in principle to the EU and Australian approach. For example, Canada has established "guiding principles for AI in government" which are based off the core values in the Digital Nations shared approach to AI.[40] This is exemplary of a trend toward multilateral efforts to regulate AI.[41] Canada has also introduced an "algorithmic assessment tool" (AIA) tool to be used by government departments and agencies to assess the risk of an AI system in government projects. The AIA is a self-assessment that is intended to "identify and assess impacts in a broad range of areas," including human rights, health, economics and environment.[42] Other efforts include the Canadian Government's "Directive on Automated Decision Making" the objective of which is *"to ensure that automated decision systems are deployed in a manner that reduces risks to clients, federal institutions and Canadian society, and leads to more efficient, accurate, consistent and interpretable decisions made pursuant to Canadian law."*[43] In the United States, no consistent framework has been adopted thus far, although standalone laws on AI exist across the country. Of note, in October 2023 President Biden issued an executive order on safe, secure and trustworthy AI, which establishes measures to "protect Americans from the potential risks of AI," including requiring the developers of the most powerful AI systems to share safety tests with the US government, developing tools to ensure safety of AI systems prior to release, protect against the use of AI to engineer dangerous biological materials, establish practices that detect AI-generated content. It also aims to protect privacy, advance civil rights and equity, stand up for patients, consumers and students, promote competition and innovation, advance American leadership abroad, and ensure responsible government use of AI.

---

[35] Australian Government Department of Industry, Science and Resources. "Australia's AI Ethics Principles," 2024. https://www.industry.gov.au/publications/australias-artificial-intelligence-ethics-framework/australias-ai-ethics-principles.
[36] Ibid.
[37] Australian Government. "National Framework for the Assurance of Artificial Intelligence in Government." 21 June 2024. https://www.finance.gov.au/sites/default/files/2024-06/National-framework-for-the-assurance-of-AI-in-government.pdf.
[38] New South Wales Government. "NSW Artificial Intelligence Assessment Framework | Digital NSW," n.d. https://www.digital.nsw.gov.au/policy/artificial-intelligence/nsw-artificial-intelligence-assessment-framework.
[39] Ibid.

**Analysis**

In this analysis I take the position that the most serious problems with the risk-based model of AI governance rest in its failure to classify, frame, and calculate risk in a way that accounts for an adequate range of possible harms that AI poses. In particular, that the classification of risk as unacceptable, high, limited or minimal in the EU AI Act is flawed because certain risks that AI poses are incalculable, and the risk-based model fails to recognise this or to provide adequate reasoning as to why certain risks are classified the way they are. As a result, the structure of the governance model does not account for the true diversity of potential risks that AI poses to all members of society. As the risk-based model of AI governance has not been fully implemented yet, the possibility for rigorous analysis is limited by lack of empirical evidence. This is not to say there is no possibility for analysis- there is substantial effort to implement the risk-based model in the AI sector internationally. However, most formal efforts are in their early stages. Because of this any analysis would not be complete unless it both pre-empted failings of the model and proposed solutions. Analysis must also recognise that no part of a governance structure is neutral and objective- every part of the model is designed by a complex set of choices,[44] which are influenced by politics, power, biases and social and economic structures of that place and time.

---

[40] Secretariat, Treasury Board of Canada. "Guiding Principles for the Use of AI in Government." Canada.ca, May 30, 2024. https://www.canada.ca/en/government/system/digital-government/digital-government-innovations/responsible-use-ai/principles.html.
[41] Marcin Szczepański, "United States approach to artificial intelligence." European Parliament. (2024) https://www.europarl.europa.eu/RegData/etudes/ATAG/2024/757605/EPRS_ATA(2024)757605_EN.pdf.
[42] Secretariat, Treasury Board of Canada. "Guiding Principles for the Use of AI in Government." Canada.ca, May 30, 2024. https://www.canada.ca/en/government/system/digital-government/digital-government-innovations/responsible-use-ai/principles.html.
[43] Secretariat, Treasury Board of Canada. "Directive on Automated Decision-Making." Canada.ca, April 25, 2023. https://www.tbs-sct.canada.ca/pol/doc-eng.aspx?id=32592.

The EU AI Act is a part of the larger trend toward risk-based regulation that has taken place over the last three decades across Western states such as Europe, Australia and America.[45] As such the proposed model carries with it established ideas about risk and risk regulation that have developed over time. Traditionally, risk is understood to be a calculation of probability x impact.[46] This is a view of risk that works well when a quantitative risk assessment is involved and under conditions in which it is possible to identify a comprehensive set of possible outcomes and the level of risk is ranked and known.[47] Qualitative risk assessments were developed to address mechanical problems in which the desired outcome was clear, and the reliability of separate components was measurable.[48] In situation where these conditions are not met, risk is not easily quantifiable- but the traditional notion of risk as probability x impact remains the most culturally accepted and widely used, even under circumstances where uncertainty is much higher. Perhaps one reason this idea remains at the forefront of risk understanding is that it is useful because it is used to legitimate regulatory decisions- Black[49] argues that the rhetoric that the government's role is to mitigate risk allows it to justify its intervention into society in the interest of that objective. This justification relies on the notion of risk as an undesirable outcome times its probability- the conventional understanding of risk. However, Black argues, the conventional idea of risk is 'culturally contestable' and creates three important possibilities for socio-political contestation that can be used to analyse the risk-based model of regulation.

The first of these points of socio-political contestation of risk is that the idea of risk as the possibility of an undesirable outcome assumes a given notion of undesirability, which is inherently normative. Any form of governance must assume a certain notion of desirability and undesirability, but the categorisation of certain risks as more important than others raises questions about more which values are most important and why- a topic that opens up possibility for debate by civil society, threatening the moral authority of governance decisions. Moreover, Black argues, the conventional idea of risk creates situations of incommensurability between risks. It essentially takes the full spectrum of human experience and categorises very different experiences as worse or better than others (when risk = probability x impact), when in reality two different situations might be very difficult to compare. For example, when comparing the environmental risk of biodiversity loss against the risk of poverty, which may be a very real decision for regulators, one comes up against two clearly 'bad' scenarios, which are not easily comparable to each other and not clearly measurable, thus the risks are 'incommensurable.' In this situation regulators may be forced to make difficult decisions and turn to their own perceptions of risk in order to make regulatory choices, which are known to often be biased based upon a person's circumstances and experiences.[50]

---

[44] Black, Julia, and Robert Baldwin. "Really responsive risk-based regulation." Law & policy 32, no. 2 (2010): 181-213. P 185.
[45] Ibid.
[46] Black, Julia, 'The Role of Risk in Regulatory Processes', in Robert Baldwin, Martin Cave, and Martin Lodge (eds), The Oxford Handbook of Regulation (2010; online edn, Oxford Academic, 2 Sept. 2010), https://doi.org/10.1093/oxfordhb/9780199560219.003.0014.
[47] WEIMER M. 'The Origins of "Risk" as an Idea and the Future of Risk Regulation.' European Journal of Risk Regulation. 2017;8(1):10-17. doi:10.1017/err.2016.14.
[48] Ibid.
[49] Black, Julia, 'The Role of Risk in Regulatory Processes', 306.

Another issue embedded in the understanding of risk used in risk-based models of governance is the idea that there is causality between a risk and what is being governed in order to affect that risk.[51] Evidently the extent to which there is a causal relationship and how confident regulators are that what they choose to do effects the outcome will impact governance decisions. In cases where there has been evidence of causality and the extent of the causality over time the regulators have an easier job- but often there simply hasn't been enough time to figure out the relationship between risks and their causes. In the case of AI regulation this issue is particularly apparent- the 'novelty' of many AI systems might make the source of certain undesirable outcomes obscure. Moreover, the importance of identifying causal relationships in the risk governance process points to a wider issue of uncertainty about the future. Uncertainty can be defined separately from risk, because while risks can be measured (at least to some extent) uncertainties cannot. The issue of uncertainty, according to Black,[52] reveals three drastically different states of knowledge about future events- known knowns, which are risks as they can be quantified, known unknowns, or uncertainties, and unknown unknowns. Weimer[53] argues that one view idea of risk that must and can be controlled is the product of an industrialised society in which technological innovation allows for greater control over nature, increased productivity and increased safety, but also for new unmeasurable and uncontrollable risks as an unintended side effect of this technological progress. Where in traditional societies hazards or dangers were seen as a given part of the world where no notion of risk existed, the idea of risk in modern industrialised societies emerges as a result of the desire for control over the future. With the progression of human industrial development, new 'manufactured risks' emerge to create a novel risk environment in which unknown risks are unknown and incalculable.[54] Thus it is possible to differentiate the notion of 'risk' from 'harm' – while risk implies probability of harm it also assumes a degree of control the future and control over the world, environment, society and technology. For some AI systems when effects have not been seen and are difficult to predict in context, I ague it is not meaningful to classify undesirable outcomes as 'risks,' rather, they are 'unknown unknowns' in the sense that both the harms they produce are unknown and the probability of those harms occurring is incalculable.

Having distinguished harms from risks, it could be said that the EU AI act and other proposed risk-based regulation models identify a range of harms that AI may cause. Indeed the classification of certain outcomes as 'unacceptable,' 'high,' 'low' and 'minimal' risk in the AI

---

[50] Black, Julia, 'The Role of Risk in Regulatory Processes,' (2010), 310.
[51] Black, Julia, 'The Role of Risk in Regulatory Processes,' (2010), 310.
[52] Black, Julia, 'The Role of Risk in Regulatory Processes,' (2010). 309.
[53] WEIMER M. The Origins of "Risk" as an Idea and the Future of Risk Regulation. European Journal of Risk Regulation. 2017;8(1):10-17. doi:10.1017/err.2016. 11.
[54] A Giddens, "Risk and Responsibility" (1999) 62 Modern Law Review 1, 3.

Act[55] assumes a notion of probability by default. It is rather unclear as to how the proposed models intend to calculate the probability of these harms occurring. Rather, by virtue of being called 'risks' they are automatically assumed to have an element of probability and impact attached to them. I argue that the issue with assuming the impact and probability of certain harms in the risk-based governance model is twofold- on one hand, there is a high degree uncertainty about what the outcomes of AI systems' use in context may be, bringing the calculability of risk into question. Moreover, the assumption of the existence of certain 'risks' without acknowledgement of how those risks were calculated (or perhaps more accurately, decided upon), and who is considered in these calculations, is deeply problematic, because of the inherent normativity that exists in any notion of risk.

As is the case with any technology, there is a degree of uncertainty about the outcomes AI will produce. Indeed what often ends up being the subject of risk regulation discussions is uncertainty rather than risk.[56] A degree of uncertainty is inherent to new technological developments, but is particularly high in AI, for example, in the case of what have been called 'frontier models.' So-called frontier models, or 'general-purpose AI systems,' are referred to by many scholars and popular technology discourse as a category of AI systems that can perform a broad range of tasks as opposed narrow models trained to do specific tasks.[57] Frontier models, as well as other powerful and new AI systems, produce uncertainty about their effects which is twofold- uncertainty about the effects themselves (what will happen, who will be affected) and uncertainty about the likelihood of those effects. For example, there may be 'dual use' scenarios in which an AI is used to intentionally cause harm by a malicious user. A prime example of a 'frontier model' is GPT-4, OpenAI's well known LLM. Such a case has famously already occurred in the case of GPT-4, although perhaps more as a deliberate illustration of uncertain outcomes. A month after GPT-4 was launched, a developer hijacked the system and ran an autonomous agent named ChaosGPT, with the aim of destroying humanity. The agent found research on nuclear weapons, recruited other AI systems and made tweets to influence users. While obviously unsuccessful in achieving destroying humanity, ChaosGPT served its purpose in illustrating some of the more 'catastrophic' risks that AI presents. Other risks that are brought up often enough in AI safety debates include the possibility of developing 'artificial general intelligence,' or so-called 'catastrophic risks' from AI such as rouge AI systems that resist being shut down, powerfully manipulate people and public belief, or get hijacked by malicious actors, are used for mass surveillance and control by authoritarian regimes.[58] These 'catastrophic' outcomes would be undeniably bad, and governance efforts should aim to decrease the possibility of them occurring. The existence of 'catastrophic risks' serves to illustrate that AI presents certain 'unknown unknowns'; harms which are unknown and possibly even inconceivable, and the probability of these harms is unknown and unquantifiable. Under these conditions, it makes little sense to call these harms 'risks,' as we have seen, the notion of risk assumes an element of probability and control over the future. Moreover, the use of risk in regulatory rhetoric is not only technically inaccurate but is also potentially harmful to the people affected by governance decisions based on risk, because risk is constructed to reflect existing power structures and used to legitimate regulatory decisions and justify regulatory intervention or non-intervention into society.

---

[55] European Union. Shaping Europe's Digital Future. "AI Act," June 26, 2024. https://digital-strategy.ec.europa.eu/en/policies/regulatory-framework-ai.
[56] Black, Julia, 'The Role of Risk in Regulatory Processes', 310.
[57] Lorraine-Tri. "Frontier AI: Heading Safely Into New Territory." Trilateral Research, July 9, 2024. https://trilateralresearch.com/emerging-technology/frontier-ai-heading-safely-into-new-territory.
[58] Centre For AI Safety. "An Overview of Catastrophic AI Risks." https://www.safe.ai/ai-risk

Black's[59] argument that the notion of risk assumes a certain normative idea of what is desirable and undesirable speaks to a wider issue with the risk-based model of AI governance- that the notions of risk, and particularly how risk is prioritised in these models (as 'unacceptable, high, limited, minimal') reflects and reproduces existing power structures. I argue that the range of 'risks' the AI act accounts for, particularly its classification of unacceptable risk systems, which include any that lead to subliminal practices, exploitative practices, social scoring systems and real-time biometric identification, is inadequate to address the full range of adverse lived experiences that may be caused by their use in a range of contexts. The range of harms that AI may potentially cause are broad, diverse, and often unknown and unexplored. But more importantly, many of the potential harms that AI systems' use in different contexts *are* known, and disproportionately affect already marginalised groups,[60] which is the subject of a large and growing body of research. Proponents of risk-based models, however, for the most part do not acknowledge the normativity inherent in their notions of risk, and instead extol the idea that these risk-based models do account for an acceptable range of harms.

Central to understanding risk regulation is understanding the questions; who gets to decide what risks are acceptable, and to whom do those risks pertain? Another way to think of this is to ask the questions: what risks are present, and what risks matter? No risk calculation can even begin to address risks if the risks that matter are not defined.[61] As Black emphasises, every aspect of a risk-based governance structure is determined by a set of individual decisions.[62] Individuals have different definitions of which risks matter, based on their culture, society, economic, political, and social status. Indeed, research has shown that there are strong psychological biases that exist in individuals' conceptions of risk and how it will affect them, dependent on factors such as their proximity to the risk, the perceived value of the risk, whether they chose to undertake it, and how much control they have over it. Experts are not exempt from these biases and sometimes show unique biases.[63] Values influence individual's conception of this, any risk calculation begins with the facts of the identity of the people who are developing it, and individuals are situated in a wider societal power structure.

---

[59] Black, Julia, 'The Role of Risk in Regulatory Processes,' (2010), 310.
[60] Shelby, Renee, Shalaleh Rismani, Kathryn Henne, AJung Moon, Negar Rostamzadeh, Paul Nicholas, N'Mah Yilla-Akbari et al. "Sociotechnical harms of algorithmic systems: Scoping a taxonomy for harm reduction." In Proceedings of the 2023 AAAI/ACM Conference on AI, Ethics, and Society, (2023) pp. 723-741.
[61] Baldwin, Robert, and Julia Black. "Driving Priorities in Risk-Based Regulation: What's the Problem?." Journal of Law and Society 43, no. 4 (2016): 569.
[62] Ibid.
[63] Black, Julia, 'The Role of Risk in Regulatory Processes', 312.

Indeed, there is a pervasive misconception in political rhetoric that risk calculations in a given risk-based model are a source of objectivity which can inform regulators on what to do. In practice, the 'science' of risk calculation is in its early stages,[64] and the idea that it a source of objectivity is far from the reality of the functioning of these structures, especially when it comes to AI. Each step of any risk calculation begins with normative judgements,[65] and by extrapolation is influenced by the greater social structure and indeed governance and power structures present at the time of its making.

For example, Borraz et. Al., writing specifically concerning food safety, argue that one reason that risk-based regulatory inspection practices vary across the EU is that the conventional definition of risk (hazard x probability of occurrence)"[66] leaves room for varying interpretations. Where France, Germany, the Netherlands and the UK are all subjected to the same regulations that mandate them to implement a risk-based approach to food safety regulations, each country uses very different risk-scoring methods that result in equally different risk evaluations. One explanation for this is that there exists no scientific consensus on how to assess risks in the food safety industry. This can be contrasted with an established body of science that addresses *hazard* research, which draws on disciplines such as chemistry, microbiology and epidemiology.[67] The difference between risk calculation and hazard calculation is reflective of the distinction between risk and harm; where harms refer to events in the world, there is little agreement on which undesirable outcomes to consider, for example, the size and characteristics of the consumer base that would be affected by non-compliant food and beverage organisations, in the context of food safety.[68] In the case of AI, these issues are magnified. The EU AI Act prioritises 'high-risk' AI systems as its main regulatory focus.[69] While this is an expected approach for any risk-based model of governance, defining what high-risk means is not a issue that the act seems to struggle with, in fact, high-risk is simply a category of AI systems that are defined as performing certain types of functions in certain contexts, such as facial recognition, inferring emotion, detecting and using people's race, gender, religion, biometric identification, etc. In the face of a lack of strong justification of these fundamental regulatory judgements, an essential consideration becomes, why did regulatory officials who developed the model define and categorise risks in the way that they did?

Analyses have already been conducted on the reasoning for the definitions of risk in the AI Act, for example, Regine, 2023[70] and Justo-Hanani, 2022.[71] An important point for the AI Act, the DISR report in Australia, and likely the risk-based model for AI regulation globally, is the political interplay between innovation and ethics concerns for regulatory bodies that makes the risk-based approach so appealing. By adopting a ban on technologies that are considered too 'high-risk' to be acceptable by the most powerful and vocal political actors, regulators can appease citizens and the international community, while adopting a laissez-faire approach to regulating AI systems that are considered low risk that allows industry actors to continue development of new products free from regulatory burden. For example, a report for Ernst and Young and Trilateral Research in 2021[72] argues that *"By adopting a proportionate approach where the complexity of regulatory compliance depends on the risk that the AI system poses, policy makers can fulfil their duty to safeguard without unduly impeding the benefits that AI can bring to society."* Such a statement exemplifies the idea that the risk-based framework, although not a perfect model, will be a happy medium, and at the same time shows that this rhetoric is one being pushed by industry itself. Such rhetoric is pushed by governments in Australia, America and the EU. For example, the DISR interim response paper[73] argues that *"A risk-based approach allows low-risk AI development and application to operate freely while targeting regulatory requirements for AI development and application with a higher risk of harm."* Drawing from a more critical approach, this justification of the use of the risk-based model can be seen as legitimating regulation decisions.[74] When risk-analysis is seen as objective, governance decisions which are based on that objective risk analysis are seen as legitimate by extension. The risk-based model is legitimated from within industry and by regulators when the narrative that it walks the line between innovation and safety is espoused. Furthermore, to use Black's assertion, the influences on decision-making about risk-based models revolves around how issues are packaged, constructed and identified. Looking outward to the broader cultural narratives about AI in western countries, ideas about risk analysis as objective relate to ideas about AI as being objective and intelligent. The name "artificial intelligence," Kate Crawford has argued in *the Atlas of AI*[75], is a fallacy- AI is neither "artificial" nor "intelligent." AI systems subsist from human labour and natural resources and the "intelligence" in question is constructed from a vast network of human decisions, labour, and ecological resources. Similarly, ideas about machine "learning" imply that AI systems have some kind of self or person that can learn, divorcing it the reality of the human decisions that created the system and thus diffusing responsibility from the harm it causes.

---

[64] Borraz, O., et al.(2022). Why regulators assess risk differently: Regulatory style, business organization, and the varied practice of risk-based food safety inspections across the EU. Regulation & Governance, 16(1), 274–292. 276.
[65] Black, Julia, 'The Role of Risk in Regulatory Processes', 312.
[66] Borraz, O., et al. "Why regulators assess risk differently: Regulatory style, business organization, and the varied practice of risk-based food safety inspections across the EU."(2022). 276.
[67] Ibid., 286.
[68] Ibid.
[69] Paul, Regine. 'European artificial intelligence "trusted throughout the world": Risk-based regulation and the fashioning of a competitive common AI market.' Regulation & Governance (2023). 2.
[70] Ibid.
[71] Justo-Hanani, Ronit. "The politics of Artificial Intelligence regulation and governance reform in the European Union." Policy Sciences 55, no. 1 (2022): 137-159.

I have proposed a criticism that the construction of the risk-based model view of risk analysis as objective, coupled with the institutional and political factors, has led to the definition of 'risk' as it is now. Thus, the classifications of 'unacceptable,' 'high' 'limited' and 'minimal' risk systems reflect conventional ideas about who and what is important and misses out on the true diversity of adverse lived experiences and outcomes that are produced by AI systems in context. In reality, the scope of potential harms from AI systems are broad and the subject of a growing body of literature. Shelby et. Al., for example, present a scoping taxonomy of sociotechnical harms from AI systems, which identifies a wide range of harms from AI systems, including representational harms that reproduce unjust social hierarchies, allocative harms relating to the inequitable distribution of resources, quality-of-service harms from AI systems that serve certain groups more poorly than others, interpersonal harms in which AI systems adversely shape relations between individuals, and social system harms.[76] Harms to marginalised groups may be a kind of experience that is completely unfamiliar to those who exist in a privileged position in society, such as different sub-types of representational harms from AI systems, like the alienation and erasure of social groups, or the denial of the opportunity to self-identify.[77] For example, in image classification systems, a person might be categorised into a group in which they do not belong without their knowledge or consent, which reduces their autonomy, ability to identify on their own terms, and control over data collection and use, and since classification systems are used in a wide range of contexts can have adverse effects on marginalised communities.[78] To use the example of 'alienation' in 'quality-of-service' harms, automatic speech recognition systems may not recognise diverse ways of speaking, and due to their race, location, or gender identity, the AI system underperforms, resulting in feelings of *"annoyance, disappointment, frustration or anger when interacting with technologies that do not recognise one's identity characteristics,"*[79] and forcing people to change how they speak and who they are, just to be able to interact with technology.[80] The stereotyping of social groups is reflected in algorithmic outputs and has adverse affects on members of certain groups, particularly those whose lives are shaped by interlocking forms of oppression,[81] - for instance, *"when a search for the term "unprofessional hairstyles" disproportionately returns images of black women."* While the AI Act, for example, classifies the risk of AI systems used in *"educational or vocational training that could determine access to education or the professional course of someone's life"* as unacceptable, it fails to acknowledge that when AI systems reproduce discriminatory social narratives, like the idea that black women's hair is 'unprofessional,' the course of peoples lives *are* affected. This, I argue, is not a 'minimal' risk but rather a serious harm that should be acknowledged by those with regulatory power and the subject of regulatory discussions which include those people who are most affected and marginalised. This is just one example of the harms that AI systems do in reality, and the range of harms is growing exponentially alongside the increased development and uptake of AI systems in different contexts.[82] The idea that categorising risk without acknowledgement of the power structures at play in the decisions behind the governance model will account for 'unacceptable' and 'high' risk applications of AI systems is flawed.

---

[72] Ezeani, G., A. Koene, R. Kumar, N. Santiago, and D. Wright. "A survey of artificial intelligence risk assessment methodologies." The global state of play and leading practices identified (2021). 4.
[73] Department of Industry Science and Resources (2024). The Australian Government's interim response to safe and responsible AI consultation, Department of Industry Science and Resources. 13.
[74] See Black, Julia, 'The Role of Risk in Regulatory Processes', 312.
[75] Crawford, Kate. The atlas of AI: Power, politics, and the planetary costs of artificial intelligence. Yale University Press, 2021.

---

[76] Shelby, Renee, Shalaleh Rismani, Kathryn Henne, AJung Moon, Negar Rostamzadeh, Paul Nicholas, N'Mah Yilla-Akbari et al. "Sociotechnical harms of algorithmic systems: Scoping a taxonomy for harm reduction." 733.
[77] Shelby, Renee, Shalaleh Rismani, Kathryn Henne, AJung Moon, Negar Rostamzadeh, Paul Nicholas, N'Mah Yilla-Akbari et al. "Sociotechnical harms of algorithmic systems: Scoping a taxonomy for harm reduction." 729.
[78] Ibid.
[79] Ibid., 730.
[80] Ibid., Table 3: Quality-of-Service harms. 731.
[81] Ibid., 728.
[82] Ibid.

# Suggestions for Future Research

There are evident drawbacks to using the risk-based model for AI governance. A comprehensive reform of the proposed policies is not within the scope of this paper and should be the subject of future research. However, as a conclusion to my analysis, I suggest that existing scholarship on governance, regulation, responsiveness and the risk-based model can inform a more effective approach to AI governance and help to recognise the complexity and scope of the harms that AI causes.

In the 2005 article *Nodal Governance,* the authors[83] argue that traditional understandings of governance as a top-down relationship between government and regulated actors is insufficient to deal with the practicality of good governance. Instead, they conceptualise governance as interconnected structures of power and change. Governance, they argue, is an organic result of the outcomes generated by a given group of people in context who are both affected by and affect upon others their beliefs about the world, their desires, and their fears, which constitutes a 'node' of governance. Nodes can take varying forms; for example, legislatures, government agencies, NGOs, firms, gangs, and neighbourhood agencies.[84] Nodes govern a system by mobilising resources, which can be social, fiscal, legal, and also regulate other nodes through networks. A governance structure, by this conception, takes into account many nodes of power with varying degrees of influence and in many forms- a network of nodes that connect to each other through various expressions of influence. Such a conception of governance "takes complexity seriously"[85] and recognises that governance constitutes a wider network of power structures acting upon each other, that influence the decisions that regulators and the subjects of regulation make. Taking this view of governance makes it necessary to acknowledge that regulatory officials, policies and laws only constitute one aspect of the many influences on a subject of regulation, and thus also provides opportunities for different forms of governance to be conceptualised and developed in broader society.

Much risk-based governance scholarship does take into account the view that governance is carried out by many varying groups in society, and even argues for an approach to risk-based governance that takes advantage of this. For example, Black[86] mentions that the risk-based approach would be most effective when used alongside other governance approaches to supplement them, and scholars of the approach to governance known as 'smart regulation' argue that regulatory tools can be most effective when utilised alongside other methods of governance. Smart regulation frames governance not as a simple interaction between government (regulator) and business (regulated entity) but sees other informal mechanisms of social control as just as, if not more, important than simple top-down governance.[87] In smart-regulation approaches, third-party entities from varying sectors, such as international organisations; trading partners; commercial institutions; peer pressure; and civil society actors, should be used in conjunction with one another to regulate an actor.[88] Smart regulation approaches emphasise using non-regulatory third-party actors to act as 'surrogate regulators,' reducing resource burden for formal regulators and allowing them to allocate resources more efficiently, just as proponents of the risk-based approach also claim the risk-based approach does. Drawing upon principles of 'responsive regulation,' the smart regulation approach also emphasises the importance and efficacy of using an 'instrument pyramid' to scale regulatory action at each level- in a similar way risk-based models aim to allocate regulatory resource consistent to levels of risk. However, smart-regulatory approaches see enforcement scaling as a process that includes the state, self-regulation, and surrogate regulators (third parties) as available options for regulation, depending on which one os most appropriate and effective, and can even be used in conjunction with one another. Where the risk-based model allocates regulatory resources according to risk level, smart regulation emphasises escalating regulatory measures up an 'enforcement pyramid' that not only considers the use of governmental resource but also the coercive power of broader social actors and the regulated entity itself.

---

[83] Burris, Scott, Peter Drahos, and Clifford Shearing. "Nodal governance." Austl. J. Leg. Phil. 30 (2005): 30.
[84] Ibid., p. 12.
[85] Ibid., 4.
[86] Black, Julia, and Robert Baldwin. "Really responsive risk-based regulation." 183.
[87] Gunningham, Niel, and David Sinclair. "Smart regulation." In Regulatory Theory: Foundations and applications. Ed. Peter Drahos, ANU Press (2017).
[88] Ibid.

Indeed, much governance scholarship emphasises the role of the broader social environment in regulation. In *Really Responsive Risk-based Regulation,* Black and Baldwin argue that the risk-based regulation model in any industry should be improved to include 'responsive regulation' practices. Black lays out five key tasks for any responsive regulatory approach, which are: (1) detecting noncompliant behaviour, (2) responding to the behaviour by developing tools and strategies, (3) enforcing those tools and strategies on the ground, (4) assessing their success or failure, and (5) modifying regulatory approaches accordingly.[89] Furthermore, in order to effectively design and execute a responsive approach to risk-based regulation, regulators should "*be attentive and responsive to five key factors: (1) the behaviour, attitudes, and cultures of regulatory actors; (2) the institutional setting of the regulatory regime; (3) the different logics of regulatory tools and strategies (and how these interact); (4) the regime's own performance over time; and, finally, (5) changes in each of these elements.*" While the scope of this paper does not allow for a comprehensive look at all of these elements, I argue considering these five key tasks for regulators would provide a good basis for further analysis and improvement of the risk-based model of AI governance.

## Conclusion

To conclude, the analysis I have presented in this paper reveals several shortcomings of the risk-based approach to AI governance that the EU AI Act and similar regulatory frameworks propose. Firstly, the risk-based model's reliance on the traditional understanding of risk as probability times impact fails to accommodate for the high degree of uncertainty that AI technologies' deployment in different contexts poses. Some AI systems, such as those that have been called 'frontier models,' introduce unprecedented levels of uncertainty that could be called 'unknown unknowns' which render conventional calculations of risk inadequate. Moreover, the inherent normativity that is present in calculations and categorisations of risk reflect and perpetuate existing power structures. The categorisation of specific AI systems as 'unacceptable,' 'high,' 'limited' and 'minimal' risk in the EU AI Act involves subjective judgements about what constitutes an acceptable level of harm and to whom. This poses challenges in addressing the broad and complex spectrum of harms that AI causes, particularly in situations where AI systems reproduce harmful societal narratives that disproportionately affect marginalised communities. Thus the current risk-based model of AI governance often fails to account for the full range of adverse lived experiences that AI systems cause. In conclusion, while the risk-based model represents a step forward in a crucial effort to establish an effective governance model for AI, further research should be conducted and should take into account alternative regulatory approaches if the risk-based model is to be effective in minimising harms that AI causes to all members of society.

---

[89] Black, Julia, and Robert Baldwin. "Really responsive risk-based regulation." Law & policy 32, no. 2 (2010): 181-213. 183.